\documentclass[10pt,a4paper,twocolumn]{article}

\usepackage[ top=1.8cm, bottom=2.0cm, left=1.7cm, right=1.7cm, columnsep=0.7cm ]{geometry}

\usepackage{amsmath,amssymb}
\usepackage{newtxtext,newtxmath}

\usepackage{graphicx} \usepackage{array} \usepackage{booktabs} \usepackage{threeparttable}

\usepackage{xcolor} \usepackage{gensymb} \usepackage[normalem]{ulem} \usepackage{microtype}

\usepackage{cite} \usepackage[hidelinks]{hyperref}

\title{ \bfseries Broadband transverse mode directional couplers\\
using partial Euler bends }

\author{ Airin Antony$^{1,*}$, Daniel Peace$^{1}$, and Jacquiline Romero$^{1}$\\[5pt] \small $^{1}$School of Mathematics and Physics, The University of Queensland, Brisbane, Australia\\ \small $^{*}$Corresponding author: \href{mailto:airin.antony@uq.edu.au} {airin.antony@uq.edu.au}
}

\begin{document}
\twocolumn[ 
\begin{@twocolumnfalse} 
\maketitle 
\begin{abstract}
Transverse modes beyond the fundamental mode are increasingly being explored to increase information capacity in both mode-division multiplexing and high-dimensional quantum information processing.
The ability to combine and separate these modes at arbitrary ratios over a broad bandwidth is central to these applications.
Motivated by the lower bending loss and intermodal crosstalk of partial Euler bends in multimode waveguides, we develop a geometric framework based on partial Euler bends for broadband power coupling between the first two transverse electric modes on a silicon-on-insulator platform.
Our methodology uses a hybrid of eigenmode expansion and finite-difference time-domain methods.
Our designs feature a mode (de)multiplexer, as well as unbalanced couplers (2/3 and 1/3 splitters).
The simulated 1-dB bandwidths of our directional couplers range from $101$ nm to $134$ nm. The best-performing fabricated mode (de)multiplexer has a measured 1-dB bandwidth of $90$ nm, a conversion efficiency of $94.0\%$ and crosstalk below $-18$ dB at $1550$ nm.
Our results validate the use of partial Euler bends for broadband transverse mode directional couplers and provide a general framework for their design. 
\end{abstract} 
\vspace{1em} 
\end{@twocolumnfalse} 
]

\section{Introduction}

The adoption of higher-order orthogonal spatial modes of the optical waveguide has shown great promise in multiple domains of integrated photonics such as optical communication and photonic interconnects \cite{invdes3, yang2022multi-interconnects1}, photonic quantum computing \cite{mohanty2017quantum}, optical processors \cite{khaled2024fully-photonic_processors}, and photonic neural networks \cite{meng2023compact-multimode_neural, sun2025highly-fanin}.
Orthogonal spatial modes (both transverse electric and transverse magnetic) can be used as independent communication channels, improving the data-carrying capacity per waveguide \cite{mojaver2024recent}.
Photonic gate operations that involve transverse spatial modes can be performed using passive linear optical devices, without the need for nonlinear effects \cite{li2019multimode}.
Mode-division-multiplexed (MDM) systems can also operate on a single laser for all channels, significantly reducing power consumption \cite{chen2014mode-singlelaser}.

Multimode couplers are key components that facilitate MDM systems; they enable power transfer from specific input to output modes at fixed splitting ratios ($\gamma$).
Broadband multimode couplers in particular, are of significant interest, as they are compatible with wavelength-division-multiplexing, which further improves the data-carrying capacity and facilitates multi-wavelength operation \cite{selective_mode_coupling}.
Many approaches have been tried to implement wavelength-insensitive multimode power coupling, each with distinct advantages and tradeoffs.
For example, multimode-interference-based couplers have been widely used for power splitting \cite{zhang2019integrated-mmi5, haines2024fabrication-mmi6, franz2021compact-mmi7, atri2025compact_mode-insensitive_case2, sun2023mode-mmi7_mode-insensitive_case, 9123578-mmi4, han2015two-mmi3, li2014compact-mmi2, uematsu2012design-mmi1}, but extending them to multiple transverse modes while maintaining compactness can be difficult \cite{li2019multimode}.
Approaches based on adiabatic couplers can provide high bandwidth and fabrication tolerance, but are also usually hundreds of micrometers in length \cite{ad_coupler_1, ad_coupler_2, ad_coupler_3}.
Inverse-designed multimode components are generally ultra-compact and ultra-broadband \cite{lime1, yang2022multi-interconnects1, invdes3, invdes4, invdes5, invdes6}; however, the inherent complexity of these approaches due to the multitude of degrees of freedom being optimized makes the final topology difficult to interpret \cite{lime1, interpretable_fab}.
There are also high-bandwidth devices that use subwavelength grating (SWG) structures \cite{swg1, swg2, swg3, swg4}, but \cite{swg_jitter} has shown that multimode SWG waveguides are extremely sensitive to fabrication jitter in the position and size of the SWG segments.
Finally, asymmetric directional couplers (DCs) can help improve bandwidth; these designs vary the widths or radii of the waveguides and have shown great promise for both single-mode \cite{circularDCref, asym1, asym2, asym3} and multimode \cite{asym4, asym5} cases.
In particular, \cite{asym2} combined straight and circular waveguides for more tunability in designing singlemode DCs with high bandwidths and arbitrary $\gamma$.
Although this method shows good promise, extending it to higher order-transverse modes can introduce additional losses and inter-mode crosstalk in the bent regions, especially if circular bends are used \cite{jiang2018low-euler_multimode1}.

Our work explores asymmetric DC designs based on partial Euler (PE) bends and straight sections for the intermodal power transfer across the first two transverse electric (TE) modes.
PE bends comprise Euler-curvature sections on either side of a circular bend.
Both Euler and PE bends are increasingly used in MDM systems due to their reduced intermodal crosstalk, improved compactness, and lower loss compared to circular bends \cite{jiang2018low-euler_multimode1, ji2022compact-euler_multimode2, cherchi2013dramatic-euler_multimode3, kita2024wide-euler_multimode4, zhang2020ultrahigh-euler_DC_racetrack}.
PE bends in particular have been shown to reduce the total bend loss by balancing the straight-to-curved waveguide transition loss with the radiative losses \cite{vogelbacher2019analysis-euler_lowloss2, fujisawa2017low-euler_lowloss3, bahadori2019universal-euler_lowloss1}.
However, despite the many advantages of PE designs for multimode waveguide bends, their use as coupled waveguides for multimode DCs remains largely unexplored.
Our work combines PE and straight geometries to generate broadband multimode DCs with arbitrary $\gamma$.

For simulating the designs, we use a combination of eigenmode expansion (EME) and  finite-difference time-domain (FDTD) methods.
EME is a frequency domain technique that simulates electromagnetic propagation through a structure by segmenting it and solving an eigenvalue problem for each segment \cite{eme1}.
EME is much faster than FDTD and has been used to simulate multiple photonic components such as tapers \cite{eme3}, circular bends \cite{eme2}, PE bends \cite{vogelbacher2019analysis-euler_lowloss2}, and for multimode waveguide design \cite{eme1}.
Here, we use EME to simulate the straight and circular coupled sections of our designs.
We chose FDTD for the coupled Euler regions and the input/output edge sections of the S-bends in our design; these geometries vary in both curvature and modal coupling along the propagation direction and FDTD helps accurately model the modal evolution through these structures.
Finally, the resulting component models are combined in Lumerical Interconnect to simulate the complete device and perform a parameter sweep over multiple straight section lengths and bend-angle values.

\section{Design structure}

Our DC designs are a combination of straight, circular, and Euler waveguide sections separated by a gap. 
An Euler curve (also known as the Clothoid curve) is a curve for which the curvature increases linearly with the geometric path length (arc length) based on a Clothoid parameter denoted by $R_{c}$.
At any point of the curve, its tangent angle $\alpha$ (measured from positive x-axis), path length $l$, and curvature $\kappa$ (inverse of radius) are related by the following equations \cite{vogelbacher2019analysis-euler_lowloss2}:

\begin{equation}
\kappa = \sqrt{2\alpha}/R_{c}
\label{eqn1a}
\end{equation}
\begin{equation}
l = R_{c}\sqrt{2\alpha}
\label{eqn1b}
\end{equation}

Equations \ref{eqn1a} and \ref{eqn1b} imply that $\kappa = 2\alpha/l$ at all points of the Euler curve. In Cartesian coordinates, the Euler curve can be constructed through scaled Fresnel integrals \cite{olver2010nist-fresnel} as follows:

\begin{equation}
x(l) = \int_0^{l} \cos{\left(\frac{t^{2}}{{2R_c^{2}}}\right)}\;dt = \int_0^{l} \cos{\left(\frac{t^{2}\alpha}{{l^{2}}}\right)}\;dt
\label{eqn1}
\end{equation}
\begin{equation}
y(l) = \int_0^{l} \sin{\left(\frac{t^{2}}{{2R_c^{2}}}\right)}\;dt = \int_0^{l} \sin{\left(\frac{t^{2}\alpha}{{l^{2}}}\right)}\;dt
\label{eqn2}
\end{equation}

\begin{figure}[t]
\centering
\includegraphics[width=\columnwidth]{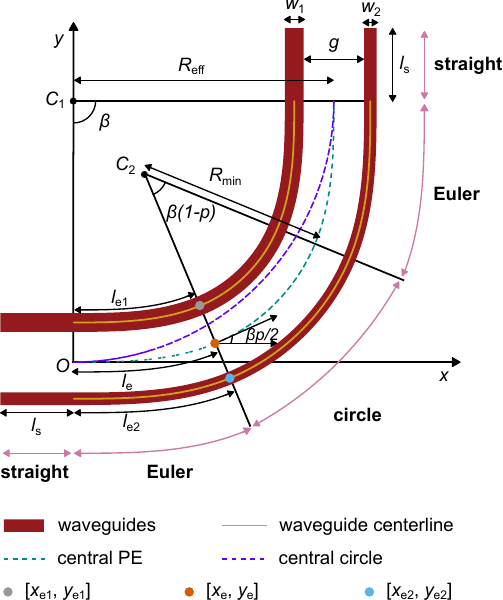}
\caption{
DC construction with combined partial Euler (PE) and straight sections.
The central PE curve is constructed based on user-specified values of $R_{eff}$ (effective radius), $\beta$ (total bend angle), and $p$ (parameter controlling the proportion of Euler and circular sections in the PE curve). The associated geometric parameters and construction points used to define the PE curve are indicated.
PE waveguides with widths $w_1$ and $w_2$, separated by a gap $g$, are constructed on either side of the central PE curve by following the corresponding waveguide centerlines.
Straight coupled sections of length $l_s$ are connected on either side of the PE waveguides.}
\label{fig1}
\end{figure}

Our PE curve is constructed by connecting Euler curves to both ends of a circular arc with matching tangent directions and radius at the intersection points.
Figure \ref{fig1} shows a PE curve at the center of both waveguides (teal dashed line), starting from the origin $O$ and constructed around the point $C_{1}$.
Let $R_{cen}$ be the Clothoid parameter of the PE, and let $\beta$ be the total bend angle.
For comparison, a circular arc with the same bend angle and end points as the central PE is included (purple dashed line).
The radius of this central circle is denoted as $R_{eff}$.
The PE curve can be divided into its Euler and circular components as shown in Figure \ref{fig1} (using pink solid lines).
The end point of the first Euler section is denoted by [$x_{e}$, $y_{e}$] (orange dot).
Let the total arc length of the first Euler section be $l_{e}$.
The adjoining circular arc has its center of curvature at $C_{2}$, a radius of $R_{min}$, and a bend angle of $\beta(1-p)$.
Here, $p\in[0,1]$ is a freely chosen parameter used to decide the proportion of Euler to the circular regions of the curve; when $p=0$, the central PE simply becomes the circle centered at $C_{1}$, or the central circle.

Next, we show that the central PE can be constructed from $R_{eff}$, $\beta$, and $p$. 
To this end, we develop the following procedure to determine $l_{e}$ for given $R_{eff}$, $\beta$, and $p$, based on three geometric observations.
First, the line segment $C_{1}C_{2}$ must have an angle $\beta/2$ with the line segment $C_{1}O$. 
Second, the tangent angle of the central PE at [$x_{e}$, $y_{e}$] must be $\beta p/2$ since the total bend angle is $\beta$.
The first two observations follow from the symmetric nature of the central PE.
Third, for a given tangent angle at [$x_{e}$, $y_{e}$], $l_{e}$ uniquely determines the coordinates of $C_{2}$ (since $l_{e}$ and the tangent angle determine [$x_{e}, y_{e}$] and $R_{min}$, based on Equations \ref{eqn1a} to \ref{eqn2}).
Therefore, for given $R_{eff}$ and tangent angle at [$x_{e}$, $y_{e}$], $l_{e}$ uniquely determines the angle between the line segments $C_{1}C_{2}$ and $C_{1}O$.
Based on these observations, it follows that $l_{e}$ is uniquely determined by $R_{eff}$, $\beta$, and $p$.
Our procedure determines $l_{e}$ by searching over a range of values through gradient descent to find the optimal Euler length at which the angle between $C_{1}C_{2}$ and $C_{1}O$ becomes $\beta/2$.
Finally, finding $l_{e}$ helps us determine [$x_{e}, y_{e}$] based on Equations \ref{eqn1} and \ref{eqn2}. Also, determining $l_{e}$ help us find the value of $R_{cen}$ using Equation \ref{eqn1b} which we use to construct the central PE based on Equations \ref{eqn1} and \ref{eqn2} and the geometric observations discussed above.

This central path is then used to construct the waveguide bends on either side.
The PE waveguide widths are denoted by $w_i$, and they are separated by a gap $g$.
Here, $i \in \{1, 2\}$ refers to the first and second waveguides, respectively.
Both waveguide centerlines (gold lines) follow the geometry of PE curves, but with a y-displacement of $\Delta y = (-1)^i~(w_{i}+g)/2$ from $O$.
The Euler section end points of the waveguide centerlines are denoted by [$x_{ei}$, $y_{ei}$] (gray and blue dots, respectively).
Our PE waveguide centerlines are constructed such that their circular sections share the same center of curvature as the circular section of the central PE (or, $C_{2}$).
Similarly, the tangent angles of the waveguide centerlines at [$x_{ei}$, $y_{ei}$] are equal to the tangent angle of the central PE at [$x_{e}$, $y_{e}$] (which equals ${\beta}p/2$).
By substituting these values in Equations \ref{eqn1} and \ref{eqn2}, the waveguide Euler section end points can be obtained from the corresponding path lengths ($l_{ei}$) as follows:

\begin{equation} \label{eq3}
x_{ei} = \int_0^{l_{ei}} \cos{\left(\frac{t^{2}p\beta}{{2(l_{ei})^{2}}}\right)}\;dt
\end{equation}

\begin{equation} \label{eq4}
y_{ei} = \int_0^{l_{ei}} \sin{\left(\frac{t^{2}p\beta}{{2(l_{ei})^{2}}}\right)}\;dt + \Delta y
\end{equation}

where, $\Delta y$ accounts for the y-displacement of the bends.
The points [$x_{ei}$, $y_{ei}$] can also be derived using displacements from [$x_{e}$, $y_{e}$] as follows:
\begin{equation} \label{eq5}
x_{ei} = x_{e} +~~ (-1)^i~\frac{w_{i}+g}{2}~\sin{\frac{\beta p}{2}}
\end{equation}

\begin{equation} \label{eq6}
y_{ei} = y_{e} -~~ (-1)^i~\frac{w_{i}+g}{2}~\cos{\frac{\beta p}{2}}
\end{equation}

The values of [$x_{e}$, $y_{e}$] were already determined while constructing the central PE for the given values of $R_{eff}$, $\beta$, and $p$.
Our procedure further determines [$x_{ei}$, $y_{ei}$] from Equations \ref{eq5} and \ref{eq6} for user-specified $w_{1}$, $w_{2}$, and $g$ values.
Then, the values of $l_{ei}$ which satisfy Equations \ref{eq3} and \ref{eq4} are determined by searching over a range of path lengths using gradient descent.
Finally, the values of $l_{ei}$ are used to derive the corresponding Clothoid parameters for the Euler sections of the waveguide centerlines (Equation \ref{eqn1b}) which are then used to construct the first Euler sections of both centerlines (Equations \ref{eqn1} and \ref{eqn2}).
Extruding along these centerlines with widths $w_{i}$ generates the coupled Euler-waveguide bends.
Next, coupled circular bends with bend angles of $\beta(1-p)$ are generated and connected to the end of the Euler section, followed by the set of mirrored Euler bends, as shown in Figure \ref{fig1}.
Finally, straight coupled waveguides with length $l_s$, widths $w_{i}$, and a gap $g$ are connected on either side to finish the construction of our combined straight and PE coupled waveguides.

\section{Simulation}

\begin{figure}[h]
\centering
\includegraphics[width=\columnwidth]{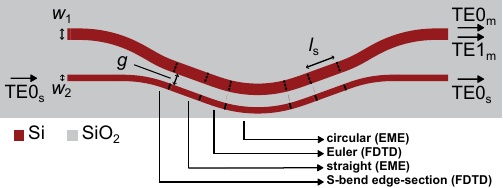}
\caption{Full DC design on a silicon-on-insulator platform partitioned into sections and their corresponding simulation methods. Parameters $w_1$, $w_2$, $g$, and $l_s$ denote the multimode waveguide width, single-mode waveguide width, coupling gap, and straight section length, respectively.
The device supports the TE0 and TE1 transverse electric modes.
TEx\textsubscript{s} and TEx\textsubscript{m} denote the TEx mode in the single-mode and multimode waveguides, respectively.}
\label{fig2}
\end{figure}

Figure \ref{fig2} shows the full DC construction designed for a standard 220-nm silicon-on-insulator platform. S-bends are added on either side for injecting the modes.
Our DC supports the TE0\textsubscript{s} mode on the single-mode waveguide, as well as the TE0\textsubscript{m} and TE1\textsubscript{m} modes on the multimode waveguide. Here, TEx\textsubscript{s} and TEx\textsubscript{m} refers to the TEx mode in the single-mode and multimode waveguides, respectively.
The DC splits the TE0\textsubscript{s} input into the TE1\textsubscript{m} and TE0\textsubscript{s} output modes, with crosstalk to TE0\textsubscript{m} as shown in Figure \ref{fig2}.
The DC is partitioned into component sections.
The Lumerical EME solver was used to obtain the S-matrices of the straight and circular coupled sections at wavelengths 1450 nm to 1650 nm with a step-size of 10 nm.
The S-matrices of the Euler-coupled sections and a short extension over the S-bends were simulated using Lumerical FDTD (wavelength range of 1450 nm to 1650 nm with auto non-uniform mesh type).
Perfectly matched layer (PML) boundary conditions were used for both EME and FDTD solvers.

We have used both EME and FDTD solvers in our model. EME is generally faster than FDTD; however, EME may exhibit reduced accuracy while trying to model geometries with abrupt variations in cross-section and may require fine segmentation and a large modal basis for geometrically complex structures \cite{eme1}. In this work, we chose FDTD for simulating the coupled Euler sections and the input/output edge sections of the S-bends.
Previously, EME has been used for modelling individual Euler bend sections \cite{vogelbacher2019analysis-euler_lowloss2} by partitioning the bend and approximating it as many circular sections with varying bend radius; however, our designs consist of two evanescently coupled Euler waveguides whose radii of curvature vary continuously along the propagation direction.
Similarly, the edge-sections of coupled S-bends have previously been approximated as simple straight coupled waveguides \cite{asym2}; however, to our knowledge, a comprehensive validation of this approximation across a range of bend radii has not been reported for coupled S-bends where the radii and the center of curvature differ for both bends.
Although EME-based approximations could be constructed for both the coupled Euler regions and S-bend edge sections by following the approach of \cite{vogelbacher2019analysis-euler_lowloss2} and \cite{asym2}, we chose FDTD to accurately capture the evolution of the guided modes through these more complex structures with simultaneously varying curvature and inter-waveguide coupling along the direction of propagation.
By using such a combination of EME for the straight and circular coupled bends and FDTD for the coupled Euler bends and the short S-bend edge-sections, we balance computational efficiency and model accuracy.

\begin{figure*}[!t]
\centering
\includegraphics[width=15cm]{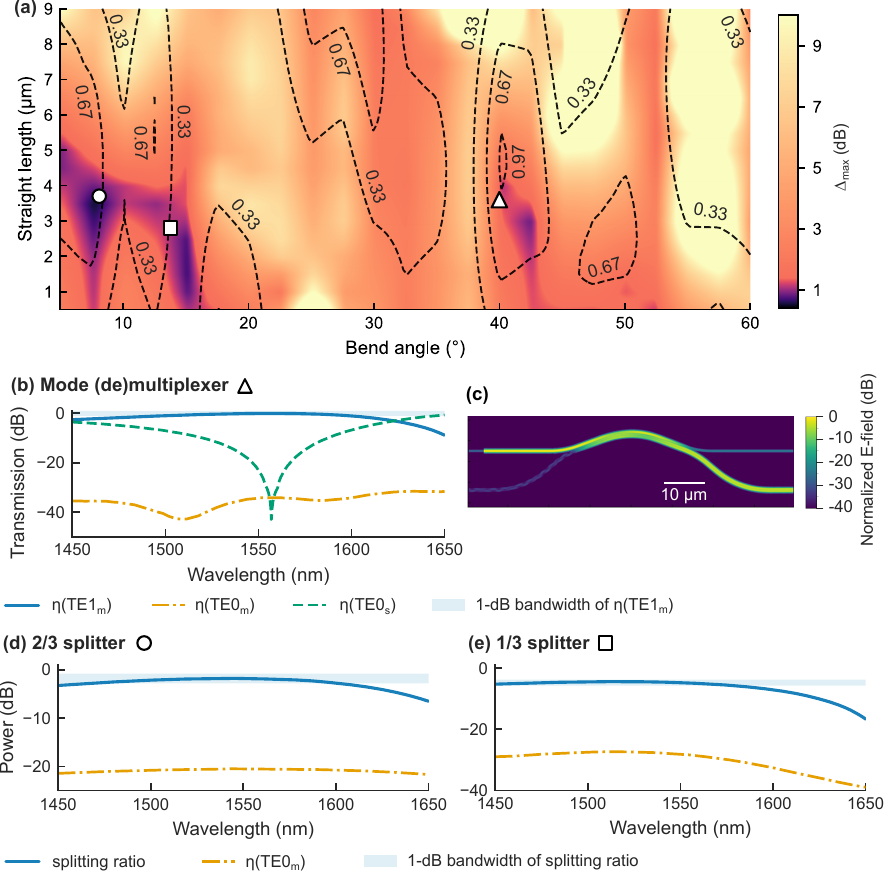}
\caption{(a) Design-space map obtained using the hybrid EME-FDTD framework and parameter sweeps in Interconnect.
The contours represent the splitting ratio at 1550 nm ($\gamma$\textsubscript{1550}), where $\gamma$ = $\eta$(TE1\textsubscript{m})/[$\eta$(TE1\textsubscript{m}) + $\eta$(TE0\textsubscript{s})].
Here, $\eta$(mode)\textsubscript{$\lambda$} denotes the conversion efficiency of the input power to the corresponding output mode at wavelength $\lambda$.
The color scale indicates $\Delta_{\rm max}$, defined as the maximum absolute deviation in $\eta$(TE1\textsubscript{m}) from its value at 1550 nm over the wavelength range of 1500 nm to 1600 nm.
The color scale is capped at $10$ dB.
The optimized design parameters are indicated using triangular (mode (de)multiplexer), circular (2/3 splitter), and square (1/3 splitter) markers.
(b) Optical spectrum of the optimized mode (de)multiplexer.
(c) Normalized electric-field (E-field) propagation through the mode (de)multiplexer.
Optical spectra of the optimized (d) 2/3 splitter, and (e) 1/3 splitter. 
}
\label{fig3}
\end{figure*}

The full DC designs were simulated using the Lumerical Interconnect software as shown in \cite{eme3}. The Interconnect solver combines the S-matrices of the constitutent sections obtained using EME or FDTD while taking into account the modal overlap at each interface.
We used the Interconnect solver to perform a parameter sweep over multiple straight section lengths and bend-angle values, hence reducing the total simulation time in our model.
For example, consider the task of simulating a number ($a\times b$) of DCs with $a$ different straight lengths and $b$ bend angles (and other parameters fixed). Rather than simulating $a\times b$ devices individually, the optical performance of all the DCs can be obtained using Interconnect from just $b$ FDTD simulations of the S-bend and Euler coupled regions, $a$ EME simulations for straight coupled sections, and $b$ EME simulations for the circular coupled bends.

We used a $p$ value of $0.3$, and an effective bend radius ($R_{eff}$) of $25~\mu$m.
The waveguide widths of $w_1$ = $0.835~\mu$m, $w_2$ = $0.4~\mu$m, and a gap of $g$ = $0.18~\mu$m were chosen for the efficient coupling of TE$0$ to TE$1$ \cite{why0.4um}.
However, any other set of values may be used; {parameters such as} the waveguide widths and gap are expected to change for a different set of transverse modes {or while trying to optimize for additional features such as fabrication tolerance}.
For our optimization, we sweep over a range of bend angles ($5 \degree$ to $60\degree$ with $2.5 \degree$ increments) and straight lengths  ($0.5~\mu$m to $9~\mu$m with $0.5~\mu$m increments).
In all simulations we use the TE0 mode of the singlemode waveguide as the source.

Our simulation results are shown in Figure \ref{fig3}(a). The contours represent the splitting ratio, $\gamma$, at 1550 nm ($\gamma$\textsubscript{1550}), where $\gamma$\textsubscript{$\lambda$} refers to the $\gamma$ at wavelength $\lambda$. $\gamma$\textsubscript{$\lambda$} is calculated as $\eta$(TE1\textsubscript{m})\textsubscript{$\lambda$} $\div$ [$\eta$(TE1\textsubscript{m})\textsubscript{$\lambda$}+$\eta$(TE0\textsubscript{s})\textsubscript{$\lambda$}], where $\eta$(mode)\textsubscript{$\lambda$} refers to the conversion efficiency of the input power to the corresponding output mode at wavelength $\lambda$.
Three $\gamma$\textsubscript{1550} values were considered, corresponding to the mode (de)multiplexer, 2/3 splitter, and 1/3 splitter designs.

Similar to the wavelength-sensitivity metric used in \cite{vogelbacher2019analysis-euler_lowloss2}, we quantified the wavelength sensitivity of our devices using the quantity $\Delta_{\rm max}$, which is the maximum absolute deviation in $\eta$(TE1\textsubscript{m})\textsubscript{$\lambda$} from its value at 1550 nm over the wavelength range of 1500 nm to 1600 nm. The metric $\Delta_{\rm max}$ is plotted as a function of the straight length and bend angle in Figure \ref{fig3}(a), and this plot serves as an initial screening tool to identify the least wavelength-sensitive regions for each splitting ratio. We used a logarithmic scale in Figure \ref{fig3}(a) to more easily identify a broadband device for each splitting ratio. We chose to plot $\Delta_{\rm max}$ in the central wavelength region between 1500 nm to 1600 nm wavelength where we could get the most uniform coupling characteristics over the entire 100 nm window.

Based on Figure \ref{fig3}(a), the least wavelength-sensitive regions were identified for each $\gamma$ and these regions were further explored via FDTD simulations in Tidy3D \cite{tidy3d}. That is, we stepped through different values of bend angles and straight-section lengths within the 2.5° and 0.5 $\mu$m region (which is the minimum step size in  Figure \ref{fig3}(a) and evaluated device-specific figures of merit. The final design parameters (obtained from Tidy3D FDTD) for the mode (de)multiplexer, 2/3 splitter, and 1/3 splitter are denoted in Figure \ref{fig3}(a) using triangular, circular, and square markers, respectively. 
For all three devices, the final parameters differed from those predicted by the hybrid EME-FDTD framework by less than 0.3° in bend angle and 0.41 $\mu$m in straight-section length, both of which are smaller than the minimum step size.
These small deviations are attributed to the simulation differences between EME and FDTD solvers, minor inaccuracies in Interconnect modal overlaps, and the unavoidable errors due to discretization. 

Figure \ref{fig3}(b) shows the spectral response of the mode (de)multiplexer, simulated using the FDTD (Tidy3D) for a wavelength range of 1450 nm to 1650 nm.
Here, bandwidth is evaluated using the variation in $\eta$(TE1\textsubscript{m}), with $\eta$(TE0\textsubscript{m}) and $\eta$(TE0\textsubscript{s}) contributing crosstalks.
Figure \ref{fig3}(c) shows the normalized electric field propagation through the mode (de)multiplexer.
Figures \ref{fig3}(d) and \ref{fig3}(e) show the spectral response of the optimized 2/3 and 1/3 splitters, respectively.
For the splitters, bandwidth is evaluated using $\gamma$ variation, with $\eta$(TE0\textsubscript{m}) reported as crosstalk.
For both splitters, the sum of $\eta$(TE1\textsubscript{m}) and $\eta$(TE0\textsubscript{s}), the quantities used to calculate $\gamma$, exceeds $-0.12$ dB ($97.3\%$) throughout the 1450 nm to 1650 nm wavelength range, indicating that only a small fraction of the input power is coupled to undesired modes or lost.

\begin{table*}[t]
  \centering
  \begin{threeparttable}
    \caption{Optical characteristics of our broadband multimode DCs}
    \label{tab:splitters}
\begin{tabular}{
>{\centering\arraybackslash}p{2.3cm}>{\centering\arraybackslash}p{0.5cm}>{\centering\arraybackslash}p{0.5cm}>{\centering\arraybackslash}p{1.0cm}>{\centering\arraybackslash}p{1.3cm}>{\centering\arraybackslash}p{0.7cm}>{\centering\arraybackslash}p{1.2cm}>{\centering\arraybackslash}p{1.0cm}>{\centering\arraybackslash}p{1.7cm}
}

\hline
Device & $l_s$ ($\mu$m) & $\beta$ ($\degree$) & Target metric (\%)& Simulated metric (\%)& IL (dB) & XT\textsubscript{m} (dB) & XT\textsubscript{s} (dB) & 1-dB bandwidth (nm) \\
\hline

Mode (de)multiplexer   & 3.6   & 40.0  & 100.00 &98.55&0.06& -34.20 & -23.06 & 101~(1499 --1600)
\\

 2/3 splitter  & 3.7 &   8.1  & 66.67   &66.51 & 0.10 & -20.43 & N/A & 134~(1467 --1601)\\
 1/3 splitter   & 2.8 & 13.8 & 33.33   & 33.47 & 0.06 & -28.07 & N/A & $>$129~($<$1450 --1579)\\

 \hline
\end{tabular}
\begin{tablenotes}\footnotesize
\item[] \textbf{Notes.}
\item [] $l_s$ and $\beta$ denote straight section length and bend angle, respectively.
\item[] Target and simulated values correspond to $\eta$(TE1\textsubscript{m})\textsubscript{$1550$} for the mode (de)multiplexer and $\gamma$\textsubscript{$1550$} for the splitter designs.
\item[] $\eta$(mode)\textsubscript{$\lambda$} is the conversion efficiency to the corresponding mode at wavelength $\lambda$.
\item[] $\gamma$\textsubscript{$\lambda$}$=$ $\eta$(TE1\textsubscript{m})\textsubscript{$\lambda$} $\div$ [$\eta$(TE1\textsubscript{m})\textsubscript{$\lambda$}+$\eta$(TE0\textsubscript{s})\textsubscript{$\lambda$}].
\item[] The desired output is TE1\textsubscript{m} for the mode (de)multiplexer and \{TE1\textsubscript{m}, TE0\textsubscript{s}\} for the splitter designs.
\item[] IL denotes insertion loss relative to the desired output mode(s) at 1550 nm.
\item[] XT\textsubscript{m/s} denotes crosstalk to TE0\textsubscript{m/s} relative to the desired output mode(s) at 1550 nm. For splitters, XT\textsubscript{s} is not applicable since TE0\textsubscript{s} is a desired output.
\item[] 1-dB bandwidth refers to the bandwidth of $\eta(\mathrm{TE1_m})$ for the mode (de)multiplexer and {$\gamma$} for the splitter designs, relative to the value at 1550 nm. The corresponding wavelength range is also reported.

    \end{tablenotes}
  \end{threeparttable}
\end{table*}

The corresponding optical performance metrics of our devices are summarized in Table \ref{tab:splitters}.
The final designs show insertion losses below $0.1$ dB and 1-dB bandwidths ranging from $101$ nm to $134$ nm.
The bandwidths of both splitters are shifted to shorter wavelengths. The 1/3 splitter exhibits a 1-dB bandwidth greater than $129$ nm, but because of this shift, its bandwidth does not fully cover the $1500$ nm to $1600$ nm wavelength range.

\subsection{Footprints and fabrication sensitivity analysis}

\begin{figure}[!b]
\centering\includegraphics[width=\columnwidth]{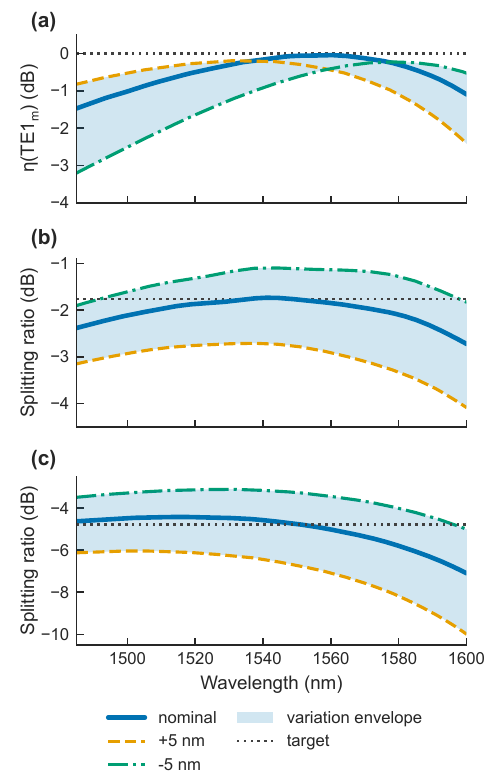}
\caption{FDTD spectra of the nominal and dilated structures for: (a) mode (de)multiplexer, (b) 2/3 splitter, (c) 1/3 splitter.
$\eta$(TE1\textsubscript{m}) is plotted for the mode (de)multiplexer, while the splitting ratio ($\gamma$) is plotted for both splitters.
A dilation of $\pm$5 nm corresponds to a $\pm$5 nm change in waveguide widths and a $\mp$5 nm variation in the coupling gap.
The shaded region denotes the variation envelope bounded by the nominal and dilated structures, while the dotted line indicates the target value.
}
\label{fig4}
\end{figure}

The primary focus of this study is the bandwidth characteristics of PE-based DC designs.
Including additional constraints such as compactness and fabrication robustness would require a much larger dataset with more parameters such as the coupling gap being optimized. This optimization is beyond the scope of our current work. Note that the experimentally tested devices consisted of input and output S-bends (the radii of which is a design choice), and a central region (PE and straight sections) where most of the coupling takes place. The footprint includes all of these regions. For the final mode (de)multiplexer, 2/3 splitter, and 1/3 splitter designs, the area of the central regions are 24.2 × 4.4 $\mu$m$^2$, 11.0 × 1.8 $\mu$m$^2$, and 11.7 × 2.0 $\mu$m$^2$, respectively, while the corresponding total footprints are 70.7 × 14.6 $\mu$m$^2$, 21.4 × 2.2 $\mu$m$^2$, and 29.3 × 3.2 $\mu$m$^2$.

Figure \ref{fig4} presents a fabrication sensitivity analysis in which our designs are subjected to a $\pm$5 nm dilation, corresponding to a $\pm$5 nm change in waveguide widths and a $\mp$5 nm variation in the coupling gap.
The results indicate that the mode (de)multiplexer is less sensitive to fabrication bias than both splitter designs at $1550$ nm.
The $\eta$(TE1\textsubscript{m}) of the mode (de)multiplexer varies by $0.57$ dB at $1550$ nm, while the $\gamma$ of the 2/3 and 1/3 splitters varies by 1.66 dB and 3.47 dB, respectively.
Based on these results, we chose the mode (de)multiplexer design for nanofabrication as a proof-of-concept for our methodology.

\section{Fabrication and measurement of the mode (de)multiplexer}

\begin{figure*}[!h]
\centering
\includegraphics[width=15cm]{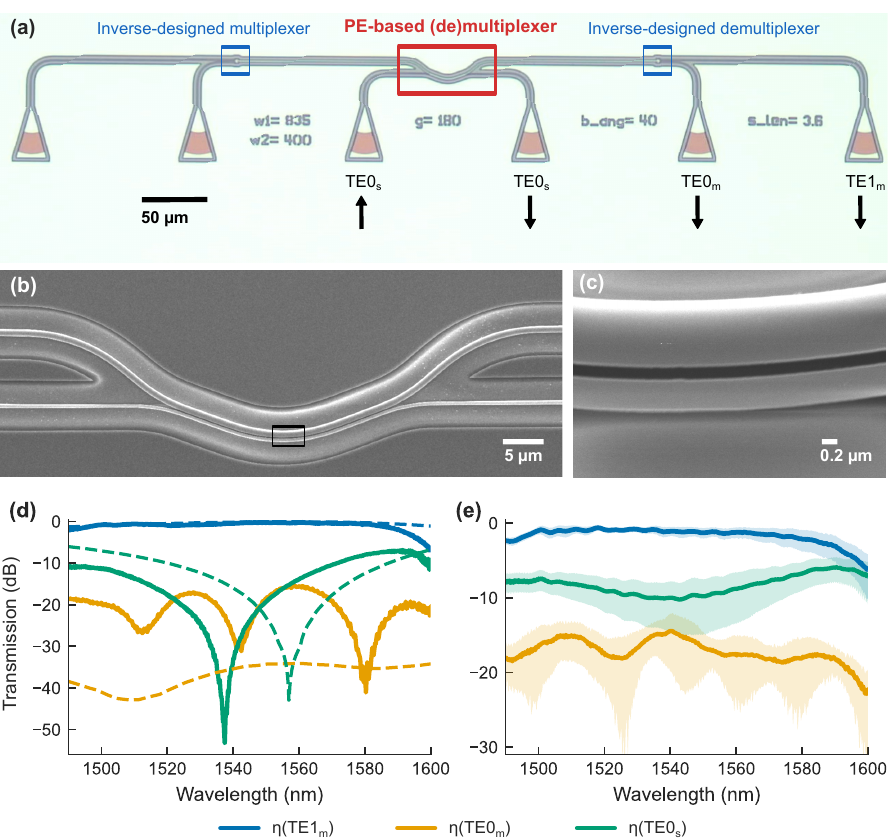}
\caption{(a) Optical microscope image of the DC circuit. (b) SEM image of the PE-based mode (de)multiplexer design; the highlighted region is magnified in (c). (c) Magnified SEM image of the PE-based mode (de)multiplexer showing the waveguides and coupling gap. (d) Measured spectra of the best-performing PE-based mode (de)multiplexer along with the corresponding simulation results (dotted lines). (e) Mean spectra of six fabricated PE-based mode (de)multiplexers, with the shaded region indicating one standard deviation.}
\label{fig5}
\end{figure*}

We fabricated six copies of our broadband mode (de)multiplexer design on a standard silicon-on-insulator platform ($220$ nm Si on $3~\mu$m SiO$2$) using electron-beam lithography and reactive ion etching.
The chip was subsequently cladded with a layer of polymethyl-methacrylate (PMMA).
Figure \ref{fig5}(a) shows the optical microscope image of the DC circuit.
The DC circuit includes shallow-etched grating couplers with an etch depth of 65 nm for coupling light between optical fibers and the fundamental TE waveguide modes.
Reference loopback measurements indicated an average peak coupling wavelength of 1558.8 nm and an average 3-dB bandwidth of 56 nm for a single grating coupler.
An inverse-designed mode (de)multiplexer  \cite{jam} is used in the circuit to separate the TE1\textsubscript{m} and TE0\textsubscript{m} modes in the multimode waveguide. 
Figure \ref{fig5}(b) shows the SEM image of the DC design, while Figure \ref{fig5}(c) provides a magnified view of the waveguides and coupling gap.

The circuits were optically characterized using a tunable semiconductor laser (Santec TSL-570) and a multiport power meter (Santec MPM-210H).
The best of six devices showed a $\eta$(TE1\textsubscript{m})\textsubscript{$1550$} of $94.0\%$ which corresponds to an insertion loss of $0.27$ dB at 1550 nm.
It also demonstrated crosstalks of $-18.33$ dB to the TE0\textsubscript{m} mode and $-19.00$ dB to the TE0\textsubscript{s} mode at 1550 nm, as well as a 1-dB bandwidth of $90$ nm in the wavelength range of 1496 nm to 1586 nm.
The measured spectra for the best device are shown in Figure \ref{fig5}(d) along with the simulation results for comparison.
Overall, the measured spectra exhibit behavior consistent with the simulations.
The best device maintains a high TE0-to-TE1 conversion efficiency, high bandwidth, low insertion loss, and crosstalk values below $-18$ dB.
The measured TE0\textsubscript{s} spectrum in Figure~\ref{fig5}(d) exhibits a shift toward shorter wavelengths relative to the simulated spectrum, and this may be due to imperfect fabrication (fabrication bias).

Figure \ref{fig5}(e) shows the mean spectral response (solid lines) of the six devices together with the corresponding standard deviation (shaded region).
The mean spectrum shows a TE0-to-TE1 coupling efficiency of $73.8\%~\pm~11.9\%$  at 1550 nm, corresponding to an insertion loss of $1.32$ dB.
The measured 1-dB bandwidth of the mean spectrum was $91$ nm, whereas the mean of the 1-dB bandwidth values across the six devices was $87\pm14$ nm.
The TE1\textsubscript{m} spectrum exhibits a larger standard deviation at longer wavelengths, which may be attributed to the finite bandwidth of the grating couplers and their increased sensitivity at wavelengths away from the grating-coupler peak wavelength.

\section{Conclusion}
We extend previous works using straight and circular waveguides for broadband single-mode DCs \cite{asym2} to multimode devices using PE bends.
Our technique uses a hybrid EME-FDTD design framework with parameter sweeps performed in Interconnect.
We designed three broadband DCs that perform TE$0$ to TE$1$ power coupling--- a mode (de)multiplexer, a 2/3 splitter, and a 1/3 splitter.
Our mode (de)multiplexer achieves a theoretical TE0-to-TE1 coupling efficiency of 98.55\% at $1550$ nm, while the unbalanced splitters deviate from the target splitting ratio by $<0.2\%$.
Their simulated 1-dB bandwidths were in the range $101$ nm to $134$ nm.
We also fabricated our mode (de)multiplexer, with the best performing device showing a TE0-to-TE1 coupling efficiency of $94.0\%$, crosstalk values $<-18$ dB, and a 1-dB bandwidth of $90$ nm.
Since our technique is based on PE bends which are well-studied for minimizing intermodal crosstalk and bend losses for multimode bends, it holds great potential for future research in higher-order-mode asymmetric DCs.

Several directions remain for future work.
First, design parameters beyond straight length and bend angle, such as coupling gap, effective radius etc., may be included in the analysis to generate a larger dataset of designs.
Second, the designs can be optimized for optical performance indicators beyond bandwidth, such as fabrication tolerance, insertion loss, compactness, and so on.
A multidimensional gradient descent or particle swarm optimization may be employed to search through a larger dataset of designs by optimizing for multiple performance characteristics simultaneously.
Finally, our work can be extended to other higher order modes using a different set of waveguide widths.
This can enable a network of multimode Euler-shaped DCs and phase shifters for broadband high-dimensional information processing on transverse modes.

\section*{Acknowledgements}
The authors thank Prof Andrew White and Charlie Hu at the University of Queensland for some help with the experiments.
The authors acknowledge the support of Microscopy Australia at the Centre for Microscopy and Microanalysis (CMM) at The University of Queensland.
This work used the Queensland node of the NCRIS-enabled Australian National Fabrication Facility (ANFF).
This work was supported by resources provided by The University of Queensland Research Computing Centre’s Bunya supercomputer (\href{https://dx.doi.org/10.48610/wf6c-qy55} {https://dx.doi.org/10.48610/wf6c-qy55}), with funding from The University of Queensland, Brisbane, Australia.

\section*{Funding}

This research was supported by the Australian Research Council Centre of Excellence for Engineered Quantum Systems (EQUS, CE170100009), and partially funded by the Australian Research Council Industrial Transformation Training Centre for Current and Emergent Quantum Technologies (IC240100012).

\section*{Author contributions}
DP and AA conceptualised the project. AA developed the methodology, designed, fabricated, characterised the devices, performed the data analysis and wrote the original draft. DP and JR supervised the project. All authors contributed to the final editing of the manuscript.

\bibliographystyle{iopart-num}
\bibliography{sample}

\end{document}